\documentclass[conference]{IEEEtran}

\usepackage{graphicx}
\usepackage{epsf}
\usepackage{amsmath, amsfonts}
\usepackage{latexsym}
\usepackage{subfigure}
\usepackage{multirow}
\usepackage{multicol}
\usepackage[table]{xcolor}
\usepackage{algorithm2e}
\usepackage[hidelinks,bookmarks=false]{hyperref}

\begin{document}

\title{\fontsize{16}{10}\selectfont Mitigation of Civilian-to-Military Interference in DSRC for Urban Operations}

\author{
Seungmo Kim and Tsigigenet Dessalgn\\
\IEEEauthorblockA{Department of Electrical and Computer Engineering\\
Georgia Southern University\\
\{seungmokim, td04611\}@georgiasouthern.edu}
}
\maketitle 
\begin{abstract}
Dedicated short-range communications (DSRC) attracts popularity in the military applications thanks to its easiness in establishment, no need for paid subscription, and wide compatibility with any other IEEE 802.11 standards. The main challenge in DSRC is congestion due to existence of only 7 channels, which may not be enough to accommodate the increased number of transmitters expected to be deployed in the near future. Recently, there are a myriad of urban operation scenarios for the military including urban warfare and humanitarian assistance/disaster relief (HA/DR). The key challenge is that the communications among the military vehicles can be interfered by civilian users. It is the desire that the messages transmitted by the military vehicles hold a higher significance so that they can avoid the interference coming from the civilian users, which is not supported in the current version of DSRC. As a remedy, this paper proposes a protocol that prioritizes the military DSRC users while muffling the civilian DSRC users. Our results show that this prioritization method achieves higher communications performances for the military DSRC users.
\end{abstract}


%%%%%%%%%%%%%%%%%%%%%%%%%%%%%%%%%%%%%%%%%%%%%%%%%
\section{Introduction}
Year 2014 marked the first year in which over 50\% of the world's population lived in an urban area, and by 2050, the proportion is projected to approach 70\% \cite{columbia18}. As these numbers climb, so does the probability that the U.S. military will be called to operate in a dense urban environment for operations ranging from disaster relief to counterinsurgency and beyond.

The recent history of U.S. military humanitarian assistance/disaster relief (HA/DR) response is indicative of the Army's likely operational tempo for the near future. From the disaster response to Haiti in 2010 to Puerto Rico in 2017, there has been no shortage of catastrophes requiring military assistance domestically and abroad.

Technologies are rapidly changing and are essential for intelligence gathering and processing (\textit{e.g.}, command, control, communications, computers, intelligence, surveillance and reconnaissance (C4ISR) applications) \cite{c4isr}. Specifically, wireless communications among the military vehicles play a pivotal role in urban operations. The reason is that urban operations pose distinct challenges to effective utilization of American military power: a high density of social and organizational ties that are difficult for outsiders to understand; complex, civilian-filled terrain that extends in three dimensions; opaque networks of formal and informal institutions; and an even more severe than usual overload of information and accompanying difficulties separating signal and noise.

\section{Related Work}
In order to maximize the performance of a dedicated short-range communications (DSRC) system among military vehicles, this paper proposes an algorithm to muffle the civilian DSRC users according to the distance to the closest military user. There is a body of previous work discussing modification of DSRC to fit in such a coexistence scenario.

To address inter-network spectrum contention between DSRC and IEEE 802.11ac, a geometry-based performance analysis framework is discussed \cite{globecom18}. Yet it investigates the impacts of external interference without proposing a solution to resolve the coexistence issue.

Adjusting the contention window (CW) size can be a solution \cite{wu2018improving}. An enhanced method is the dynamic control backoff algorithm (DCBTA) model in which the number of transmitter stations has a direct impact on the performance of the system \cite{alkadeki2015performance}. On the other hand, a distance-based routing protocol's performance is better for traffic load environment of vehicular ad-hoc networks (VANETs) \cite{ramakrishna2012dbr}.

Basic safety messages (BSMs) are broadcast and periodic. Using these unique characteristics of a BSM, the backoff mechanism of was modified \cite{cps_5}. The method uses the expiration of periodic safety messages in order to decide the value of CW. The method helped to increase the reception probability. However, it considers limited lifetime of cooperative awareness message (CAM). A CAM can be transmitted only when back off time is zero. Moreover, there is a probability of the message to expire before transmission \cite{cps_5}.

The authors of this present paper proposed a method to reduce the channel congestion based on the inter-vehicle distance in a DSRC system \cite{southeastcon19}. However, it is not directly applicable to mitigation of military-to-civilian interference.

Distinguished from the aforementioned limitations in the current literature, the paramount contributions of this paper can be identified as follows.
\begin{enumerate}
\item It is the first work that discusses the coexistence and interplay between military and civilian vehicles based on an IEEE 802.11-based system.
\item It provides a comprehensive analysis framework to measure for the civilian-to-military interference.
\begin{itemize}
\item It models the \textit{the distance from a civilian user to the closest military transmitter} as the key criterion that enables the proposed protocol.
\item It provides a generalized framework that models a mixture of the military and civilian vehicles via a two-tier Poisson point process (PPP).
\end{itemize}
\end{enumerate}

%%%%%%%%%%%%%%%%%%%%%%%%%%%%%%%%%%%%%%%%%%%%%%%%%%
\section{System Model}
\subsection{Geometry}
\begin{figure}[hbtp]
\centering
\includegraphics[width = 0.8\linewidth]{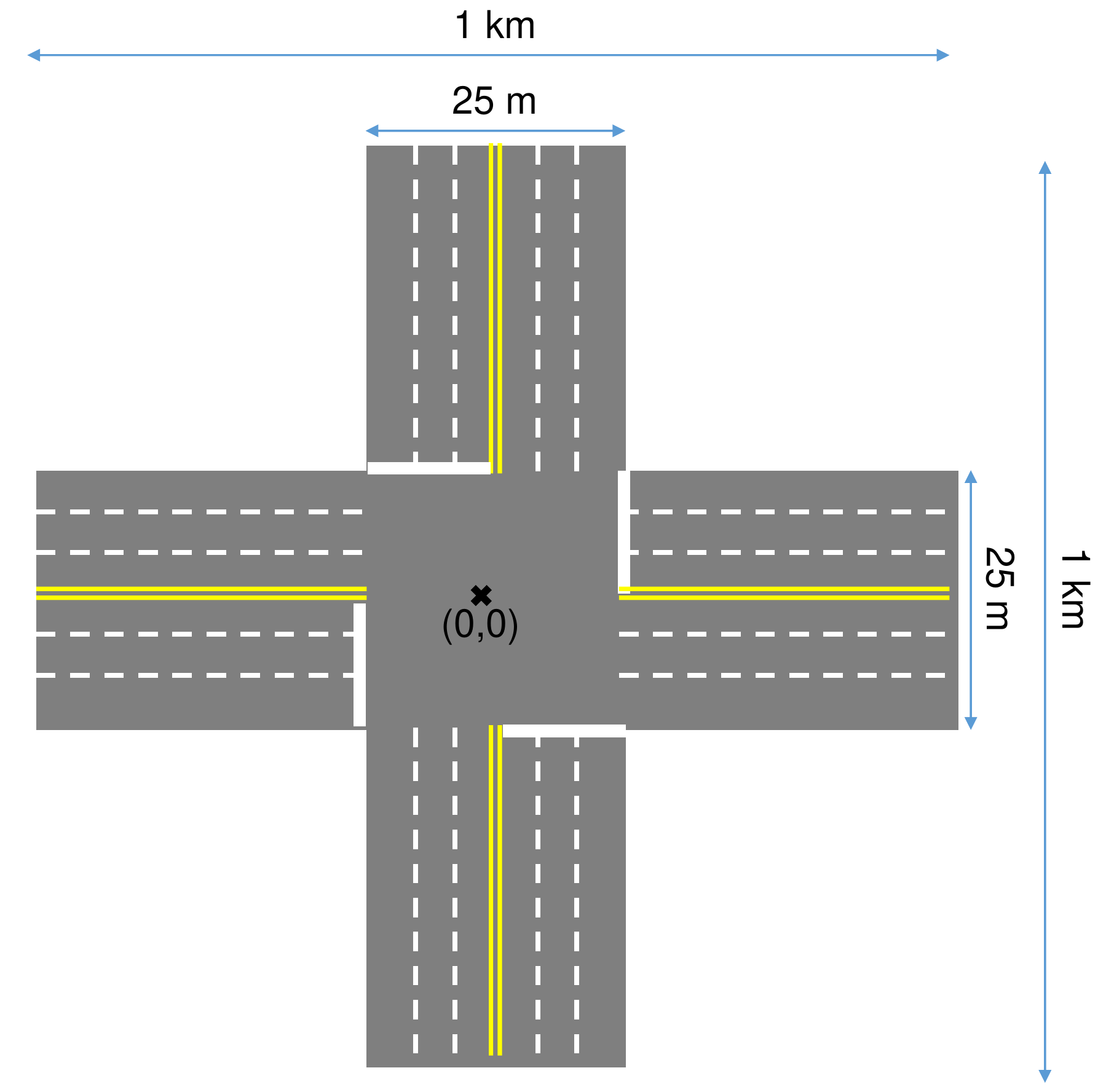}
\caption{$\mathbb{R}^2$: A junction of two 6-lane road segments}
\label{fig_system_model}
\end{figure}

A two-dimensional road segment $\mathbb{R}^2$ is defined with a 4-way junction of two 6-lane road segments, as illustrated in Fig. \ref{fig_system_model}. The position of a node is denoted as $\mathbf{x}$$=$$(x, y)$. In reference to the origin, $\mathcal{O}$$=$$(0,0)$, defined at the very center of the junction, lanes are defined at $y = \{-19, -13, -7, 7, 13, 19\} \forall x$ for the `horizontal' road segment and $x=\{-19, -13, -7, 7, 13, 19\} \forall y$ for the `vertical' road segment. No `lane changing' is assumed, considering the short length of a road segment. Also, we assume that once a vehicle reaches the end of the road segment, it wraps around to the other end of the road. This is of paramount importance to maintain fixed vehicle density and, hence, same levels of interference throughout the simulations.

The distribution of the nodes follows Poisson point process (PPP). This paper considers a two-tier PPP where the military vehicles and civilian vehicles are distributed in $\mathbb{R}^2$ according to independent homogeneous PPPs $\Phi_{\text{m}}$ and $\Phi_{c}$ with densities $\lambda_{\text{m}}$ and $\lambda_{c}$, respectively.

\subsection{Communications}
Assume that the vehicles (both military and civilian) distributed in $\mathbb{R}^2$ is fully connected: every vehicle is supposed to be equipped with communication functionality. As such, each vehicle is able to broadcast a BSM every 100 msec--\textit{i.e.}, 10 Hz of the BSM broadcast rate.

Distributed coordination function (DCF) is defined as the basic access mechanism of IEEE 802.11, which forms the basis for the physical (PHY) and medium access control (MAC) layers of DSRC \cite{cheng2014novel}. DCF is regularly used in vehicular communication in a contention-based manner \cite{wu2018improving}. Also, note that the DCF is assumed to operate in a saturated-throughput scenario \cite{bianchi2000performance}. The purpose of this is to analyze the military vehicles' communications in a worst-case scenario, which can provide a conservative guideline for the military DSRC users deployed in urban operations.

In a random 802.11 DCF, a station first listens to the activity of the channel in order to transmit a new packet. If the channel is idle for the period of a distributed inter-frame space (DIFS), the station transmits. However, if the channel is busy during the DIFS time, the station generates random backoff time before transmitting the packet. The time value is represented by contention window (CW) size and the transmitter station selects a random backoff time counter as $k\times T_{slot}$ where $k$ is a random integer within $[0, \text{CW}]$ with CW $\in [\text{CW}_{\text{min}}, \text{CW}_{\text{max}}]$. As shall be discussed in Section \ref{sec_proposed}, we propose a protocol that controls the value of $k$ to prioritize transmission from a military vehicle over that from a civilian vehicle.

\begin{table}[t]
\caption{Summary of key abbreviation and notation}
\centering
\begin{tabular}{p{0.53 in} p{2.5 in}}
\hline
\textbf{Abbreviation} & \textbf{Description}\\
\hline
BSM & Basic safety message\\
CSMA & Carrier-sence multiple access\\
CW & Contention window\\
DLVY & Packet successful delivery\\
EXP & Packet expiration\\
HN & Packet collision due to hidden node\\
PDR & Packet delivery rate\\
PPP & Poisson point process\\
SYNC & Packet collision due to synchronized transmission\\
\hline\hline
\textbf{Notation} & \textbf{Description}\\
\hline
$\mathsf{d}\left(\mathtt{x}_{i}, \mathtt{x}_{j}\right)$ & Distance between nodes $i$ and $j$\\
$\mathsf{d}_{\text{TH}}$ & Threshold for military-civilian distance\\
$l_{\text{bsm}}$ & Length of a BSM in slots\\
$L_{\text{bsm}}$ & Length of an inter-BSM interval d in slots\\
$\lambda_{\left(\cdot\right)}$ & Density (the number of points) of a PPP\\
$N_{pkt}$ & Number of packets\\
$N_{sb}$ & Number of ``busy'' slots\\
$N_{sc}$ & Number of ``collided'' slots\\
$n_{cs}$ & Number of nodes in vehicle $\mathsf{vR}$'s $r_{cs}$\\
$n_{col}$ & Number of nodes involved in a ``collision''\\
$\mathbb{R}^2$ & Space formed by two 6-lane road segments (See Fig. \ref{fig_system_model})\\
$r_{cs}$ & A node's carrier sense range\\
$r_{tx}$ & A node's transmission range\\
$\mathtt{x} =\left(x,y\right)$ & Position of a node \\
$\mathsf{vR}, \mathsf{vT}, \mathsf{vC}$ & Receiver, Transmitter, Colliding transmitter vehicle\\
\hline
\end{tabular}
\label{table_notation}
\end{table}

%%%%%%%%%%%%%%%%%%%%%%%%%%%%%%%%%%%%%%%%%%%%%%%
\section{Analysis}\label{sec_analysis}
\subsection{Result of a Packet Transmission}\label{sec_analysis_packet}
When a BSM is generated, four types of result are possible: successful packet delivery, packet expiration, collision due to synchronized transmission, and collision due to a hidden node \cite{cps_5}\cite{globecom18}. Suppose that an inter-BSM interval consists of $L_{\text{bsm}}$ slots. Then, one can compute the probability that vehicle $\mathsf{vR}$ senses $N_{sb}$ slots to be busy, namely \textit{busy slot probability} measured at $\mathsf{vR}$, which is given by
\begin{align}\label{eq_pb}
\hspace{-0.1 in}
\mathsf{P}_b &=\frac{\mathbb{E}\left[N_{sb}\right]}{L_{\text{bsm}}}\nonumber\\
&= \frac{\left(\lambda_{m} + \lambda_{c}\right) \pi r_{cs}^2 l_{\text{bsm}}}{\left|\mathbb{R}^2\right|L_{\text{bsm}}} \left( 1 - \mathsf{P}_{exp} - \frac{\mathsf{P}_{sync}}{2} - \frac{\mathsf{P}_{hn}}{4} \right).
\end{align}
Refer to Table \ref{table_notation} for the notations used in Eq. (\ref{eq_pb}).

In order to start transmission of a BSM in the $k$th slot in a beaconing period, the transmitter vehicle $\mathsf{vT}$ must not experience a beacon expiration before the time slot. Then, the first slot is uniformly chosen among the $L_{\text{bsm}}$ slots of the beaconing period. As such, the probability of vehicle $\mathsf{vT}$ starting a transmission in the slot can be given by
\begin{align}\label{eq_ptr}
\mathsf{P}_{tx} &= \frac{1 - \mathsf{P}_{exp}}{L_{\text{bsm}}}.
\end{align}
Note that $\mathsf{P}_{tx}$ will act as a key parameter for definition of the ``collision'' probabilities, $\mathsf{P}_{sync}$ and $\mathsf{P}_{hn}$, in this subsection.

\subsubsection{Beacon Expiration (EXP)}\label{sec_analysis_wo_Pexp}
To experience a packet expiration, a vehicle first finds the channel busy when a BSM is received from the network layer. This triggers a backoff of $B$, and the condition for the BSM to expire is that the node senses less than $B$ idle slots in the next beaconing period, which is formulated as
\begin{align}\label{eq_pidle}
\mathsf{P}_{idle} \left(B\right) = \displaystyle \sum_{j=0}^{B-1} \left(\begin{matrix}L_{\text{bsm}} \\ j\end{matrix}\right) \left(1-\mathsf{P}_{b}\right)^{j} \mathsf{P}_{b}^{L_{\text{bsm}}-j}.
\end{align}
It leads to the probability of an EXP as \cite{cps_5}
\begin{align}\label{eq_pexp}
\mathsf{P}_{exp} = \frac{\mathsf{P}_{b}}{\text{CW}} \displaystyle \sum_{B=1}^{\text{CW}} \mathsf{P}_{idle} \left(B\right).
\end{align}
Note that a backoff is assumed to be uniformly chosen within $\left[0, \text{CW}-1\right]$.

\subsubsection{Synchronized Transmission (SYNC)}\label{sec_analysis_wo_Psync}
Now we consider collisions among DSRC vehicles, starting from the first type of collision, SYNC. We calculate the probability of ``no SYNC'' that interferes a transmission between vehicles $\mathsf{vR}$ and $\mathsf{vT}$, which is denoted by $\mathsf{P}_{\sim sync}$ and given in (\ref{eq_pnoS}). For that, we need two probabilities. First, we need to know the probability that $n_0$ vehicles exist in $\mathcal{S}_{\mathsf{T} \cap \mathsf{R}}$, which is given by
\begin{align}\label{eq_pn0}
\mathsf{P}_{n_0} = \frac{1}{|\mathbb{R}^2|^2} \int_{\mathtt{x}_{\mathsf{T}} \in \mathbb{R}^2} \int_{\mathtt{x}_{\mathsf{R}} \in \mathbb{R}^2} \mathbb{P}\big[ \mathbb{N}\left[\mathcal{S}_{\mathsf{T} \cap \mathsf{R}}\right] = n_0 \big] d\mathtt{x}_{\mathsf{T}}d\mathtt{x}_{\mathsf{R}}
\end{align}
where
\begin{align}\label{eq_pn0_where}
&\mathbb{P}\bigl[ \mathbb{N}\left[\mathcal{S}_{\mathsf{T} \cap \mathsf{R}}\right] = n_0 \bigr]\nonumber\\
&{\rm{~~~~~}}= \left(\begin{matrix}n_{cs} - 1 \\n_0\end{matrix}\right) \left(1-\mathsf{P}_{v \in \mathcal{S}_{\mathsf{T} \cap \mathsf{R}}}\right)^{n_{cs} - n_0 - 1} (\mathsf{P}_{v \in \mathcal{S}_{\mathsf{T} \cap \mathsf{R}}})^{n_0}
\end{align}
and
\begin{align}\label{eq_Scol_sync}
\hspace{-0.1 in}
\mathcal{S}_{\mathsf{T} \cap \mathsf{R}} = \bigl\{ \mathcal{S}_{r_{cs}}\left(\mathsf{vT}\right) \cap \mathcal{S}_{r_{cs}}\left(\mathsf{vR}\right) | {\rm{~~}} 0 \le \mathsf{d}\left( \mathtt{x}_{\mathsf{T}},\mathtt{x}_{\mathsf{R}} \right) \le r_{tx} \bigr\}
\end{align}
with $\mathsf{d}\left( \mathtt{x}_{\mathsf{T}},\mathtt{x}_{\mathsf{R}} \right)$ and $r_{tx}$ being a random variable and a constant, respectively.

Note that the entire number of ``neighbors'' is regarded to be $n_{cs} - 1$, excluding $\mathsf{vT}$. Also, suppose that vehicle $v$, a neighbor of vehicle $\mathsf{vR}$, can exist as close to as $0$ and as far from as $r_{cs}$ of $\mathsf{vR}$. The probability that $v$, a neighbor of $\mathsf{vR}$, belongs to Set $\mathcal{S}_{\mathsf{T} \cap \mathsf{R}}$ is expressed as
\begin{align}\label{eq_pSsync}
\mathsf{P}_{v \in \mathcal{S}_{\mathsf{T} \cap \mathsf{R}}} = \mathbb{P}\left[v \in \mathcal{S}_{\mathsf{T} \cap \mathsf{R}} \right] = \frac{\mathsf{A}_{\mathsf{T} \cap \mathsf{R}}}{\pi r_{cs}^2}.
\end{align}
where $\mathsf{A}_{\mathsf{T} \cap \mathsf{R}}$ denotes the area of the intersection of $r_{cs}$'s of vehicles $\mathsf{vT}$ and $\mathsf{vR}$, and $\mathtt{x}_{\mathsf{R}}$ denotes position of $\mathsf{vR}$.

Second, with $n_0$ nodes in Set $\mathcal{S}_{\mathsf{T} \cap \mathsf{R}}$ from (\ref{eq_pn0}), we define the probability that $\mathsf{vT}$ is the only transmitter among the $n_0$ vehicles neighboring $\mathsf{vR}$. In other words, it is the probability that vehicle $\mathsf{vT}$ transmits a beacon without interference from any of the $n_0$ neighbors of vehicle $\mathsf{vR}$. Assume that every vehicle in $\mathbf{R}_{sys}^2$ has the same $r_{cs}$. Since every vehicle has the same $r_{cs}$, the vehicles $\mathsf{vR}$ and $\mathsf{vT}$ have $n_0$ neighbors in common. Now, we can define the probability as \cite{cps_5}
\begin{align}\label{eq_pnoSn0}
\mathsf{P}_{\sim sync|n_0} = \displaystyle \sum_{n_0 = 0}^{n_{cs}-1} \mathbb{P}\big[\text{No SYNC} \text{ } | \text{ } \mathbb{N}\left[\mathcal{S}_{\mathsf{T} \cap \mathsf{R}}\right] = n_0 \big]
\end{align}
where
\begin{align}\label{eq_pnoSn0_where}
\hspace{-0.1 in}
\mathbb{P}\big[\text{No SYNC} \text{ } | \text{ } \mathbb{N}\left[\mathcal{S}_{\mathsf{T} \cap \mathsf{R}}\right] = n_0 \big] &= \displaystyle \sum_{l = 0}^{L_{\text{bsm}}-1} \mathsf{P}_{tx} \left(1-\mathsf{P}_{tx}\right)^{n_0}\nonumber\\
&= L_{\text{bsm}} \mathsf{P}_{tx} \left(1-\mathsf{P}_{tx}\right)^{n_0}.
\end{align}

By using the probabilities (\ref{eq_pn0}) and (\ref{eq_pnoSn0}), we can generalize the number of neighbors of vehicle $\mathsf{vR}$ as $0 \le \mathbb{N}\left[\mathcal{S}_{\mathsf{T} \cap \mathsf{R}}\right] \le n_{cs}-1$. Note that $\text{max } \mathbb{N}\left[\mathcal{S}_{\mathsf{T} \cap \mathsf{R}}\right] = n_{cs}-1$ instead of $n_{cs}-1$ because among the total $n_{cs}$ vehicles neighboring $\mathsf{vR}$, the desired transmitter vehicle $\mathsf{vT}$ should transmit. With the generalized $\mathbb{N}\left[\mathcal{S}_{\mathsf{T} \cap \mathsf{R}}\right]$, the probability of ``no SYNC'' between vehicles $\mathsf{vR}$ and $\mathsf{vT}$ can be obtained as
\begin{align}\label{eq_pnoS}
\mathsf{P}_{\sim sync} = \frac{\mathsf{P}_{\sim sync|n_0}}{|\mathbb{R}^2|^2} \int_{\mathtt{x}_{\mathsf{T}} \in \mathbb{R}^2} \int_{\mathtt{x}_{\mathsf{R}} \in \mathbb{R}^2} \mathsf{P}_{n_0} d\mathtt{x}_{\mathsf{T}}d\mathtt{x}_{\mathsf{R}}\end{align}
which is performed via simulations.

Finally, the probability of a SYNC can be calculated as
\begin{align}\label{eq_psync}
\mathsf{P}_{sync} = 1 - \mathsf{P}_{\sim sync}.
\end{align}

\subsubsection{Hidden-Node Collision (HN)}\label{sec_analysis_wo_Phn}
A HN is defined as a collision caused by a ``colliding'' transmitter that is located outside of the receiver vehicle $\mathsf{vR}$`s $r_{cs}$. Specifically, the ``vehicles having the potential to collide'' are located inside of vehicle $\mathsf{vR}$'s $r_{cs}$ but outside of $\mathsf{vT}$'s $r_{cs}$. One can formulate this in a similar manner as in (\ref{eq_Scol_sync}); recall that $\mathcal{S}_{\cdot}$ denotes a set of vehicles positioned in a certain vehicle's $r_{cs}$. Therefore, for HN, one defines $\mathcal{S}_{\mathsf{R} - \mathsf{T}}$, which is given by
\begin{align}\label{eq_Scol_hn}
\hspace{-0.14 in}
\mathcal{S}_{\mathsf{R} - \mathsf{T}} = \bigl\{ \mathcal{S}_{r_{cs}}\left(\mathsf{vR}\right) - \mathcal{S}_{r_{cs}}\left(\mathsf{vT}\right) | {\rm{~~}} 0 \le \mathsf{d}\left(\mathtt{x}_{\mathsf{R}},\mathtt{x}_{\mathsf{T}}\right) \le r_{tx} \bigr\}.
\end{align}
Note that as in (\ref{eq_Scol_sync}), $0 \le \mathsf{d}\left( \mathtt{x}_{\mathsf{T}},\mathtt{x}_{\mathsf{R}} \right) \le r_{tx}$ because vehicles $\mathsf{vR}$ and $\mathsf{vT}$ are in each other's $r_{tx}$. Also, as in (\ref{eq_pSsync}), the probability that a vehicle $v \in \mathcal{S}_{r_{cs}}$ being an element of Set $\mathcal{S}_{\mathsf{R} - \mathsf{T}}$ in (\ref{eq_Scol_hn}) is expressed as
\begin{align}\label{eq_pShn}
\mathsf{P}_{v \in \mathcal{S}_{\mathsf{R} - \mathsf{T}}}
&= \frac{\mathsf{A}_{\mathsf{T} - \mathsf{R}}}{\pi r_{cs}^2}
\end{align}
where $\mathsf{A}_{\mathsf{T} - \mathsf{R}}$ denotes the area of the complement set of $\mathsf{A}_{\mathsf{T} \cap \mathsf{R}}$ within the $\pi r_{cs}^2$ of $\mathsf{vR}$.

Packet behavior in time slots is another significant difference between SYNC and HN. While a SYNC only occurs when two colliding transmissions concide at a certain time slot, a HN can occur even when the starting points of the colliding transmissions are not lined up at the same time instant. Specifically, vehicle $\mathsf{vR}$ is defined to experience a HN not only (i) when the colliding vehicle $\mathsf{vC}$ starts its transmission at the same time slot with $\mathsf{vR}$, but also (ii) when a transmission that $\mathsf{vC}$ started before still remains in effect upon the time of $\mathsf{vR}$'s transmission. It is possible because $\mathsf{vC}$ cannot be sensed by $\mathsf{vR}$ due to being located outside of $\mathsf{vR}$'s $r_{cs}$.

To formally write this relationship, a HN occurs at the receiver vehicle $\mathsf{vR}$ when any $n_0 \ge 1$ nodes in $\mathcal{S}_{\mathsf{R} - \mathsf{T}}$ either (i) starts to send or (ii) is already sending a beacon. In other words, the timing can be identified as (i) any of the $l_{\text{bsm}}$ slots, or (ii) any of the preceding $l_{\text{bsm}}-1$ slots that are occupied by vehicle $\mathsf{vT}$, respectively. The probability that neither of the two occurs is formulated as
\begin{align}\label{eq_PnoHn0}
\mathsf{P}_{\sim hn|n_0} = \displaystyle \sum_{n_0 = 0}^{n_{cs}-1} \mathbb{P}\biggl[\text{No HN} \text{ } | \text{ } \mathbb{N}\left[\mathcal{S}_{\mathsf{R} - \mathsf{T}}\right] = n_0 \biggr]
\end{align}
where
\begin{align}\label{eq_PnoHn0_where}
&\mathbb{P}\biggl[\text{No HN} \text{ } | \text{ } \mathbb{N}\left[\mathcal{S}_{\mathsf{R} - \mathsf{T}}\right] = n_0 \biggr]\nonumber\\
&{\rm{~~~~~~~~~~~~~~~}}= \displaystyle \sum_{n_0 = 0}^{L_{\text{bsm}}-1} \mathsf{P}_{tx} \left(1 - \mathsf{P}_{tx}\right)^{n_0 \left(2l_{\text{bsm}} - 1\right)}\nonumber\\
&{\rm{~~~~~~~~~~~~~~~}} = L_{\text{bsm}} \mathsf{P}_{tx} \left(1 - \mathsf{P}_{tx}\right)^{n_0 \left(2l_{\text{bsm}} - 1\right)}.
\end{align}

As a consequence, by using $\mathsf{P}_{n_0}$ from (\ref{eq_pn0}), the probability that vehicle $\mathsf{vR}$ experiences no HN can be obtained as
\begin{align}\label{eq_pnoH}
\mathsf{P}_{\sim hn} = \frac{\mathsf{P}_{\sim hn|n_0}}{|\mathbb{R}^2|^2} \int_{\mathtt{x}_{\mathsf{T}} \in \mathbb{R}^2} \int_{\mathtt{x}_{\mathsf{R}} \in \mathbb{R}^2} \mathsf{P}_{n_0} d\mathtt{x}_{\mathsf{T}}d\mathtt{x}_{\mathsf{R}}.
\end{align}

Finally, the probability of a HN can be calculated as
\begin{align}\label{eq_phn}
\mathsf{P}_{hn} = 1 - \mathsf{P}_{\sim hn}.
\end{align}

\subsubsection{Successful Delivery (DLVY)}\label{sec_analysis_wo_Psuc}
Once a packet is ``transmitted'' (not expired at a transmitter), it is assumed that every receiver not going through any of SYNC and HN is able to successfully receive the packet. Therefore, the probability of a successful packet reception is formulated as
\begin{align}\label{eq_ps}
\mathsf{P}_{suc} = \left( 1 - \mathsf{P}_{exp} \right) \left( 1 - \mathsf{P}_{sync} \right)\left( 1 - \mathsf{P}_{hn} \right).
\end{align}

\subsection{Metrics}\label{sec_analysis_metrics}
This paper uses the following three popular metrics to evaluate the performance of the DSRC system wehrein civilian and military vehicles coexist.

\subsubsection{Packet Delivery Rate}
The packet delivery rate (PDR) is defined as the rate at which a packet is successfully delivered to a desired destination node compared to the number of all packets sent. In a DSRC system setting, we assume that $\mathsf{P}_{suc}$ represents a PDR. This is reasonable since $\mathsf{P}_{suc}$ is defined as the probability that a BSM does not undergo any of EXP, SYNC, and HN.

\subsubsection{Average Delay}
Considering the significance of real-time operations in safety-critical applications, analysis on delay is essential for evaluating the performance of a DSRC system. Considering the four types of packet result, this paper defines the \textit{average delay} as
\begin{equation}\label{eq_Ttot}
T_{tot} =\mathsf{P}_{exp}T_{w} + \mathsf{P}_{suc}T_{s} + (\mathsf{P}_{ch}+\mathsf{P}_{cs})T_{c}.
\end{equation}
Note that $T_{s}$ and $T_{c}$ denote the times taken for a successful delivery and that for a packet collision, respectively, which are given by
\begin{align}\label{eq18}
T_{c} &= \text{H}+T_{pl}+\text{DIFS}+\delta\\
T_{s} &= \text{H}+T_{pl}+\text{SIFS}+\delta+\text{ACK}+\text{DIFS}+\delta
\end{align}
where H denotes the length of a header. Lastly, the average backoff time is $T_{w}$ formulated as
\begin{equation}\label{eq31}
T_{w}=\frac{\text{CW}_{\text{min}} T_{slot}}{2}
\end{equation}
where $T_{slot}$ is a slot time and $\text{CW}_{\text{min}}$ is minimum backoff window size \cite{park2014coexistence}\cite{vinel20123gpp}.

\subsubsection{Throughput}
As another metric to measure the performance of a DSRC system, this paper also defines the throughput averaged over all possible four types of result for a packet, which is formulated as
\begin{equation}\label{eq16}
S=\frac{\mathsf{P}_{suc}T_{pl}}{T_{tot}}
\end{equation}
where $T_{pl}$ represents an average time length of a payload.

\begin{figure*}[t]
\vspace{-0.2 in}
\centering
\minipage{0.35\textwidth}
\hspace{-0.2 in}
\includegraphics[width=\linewidth]{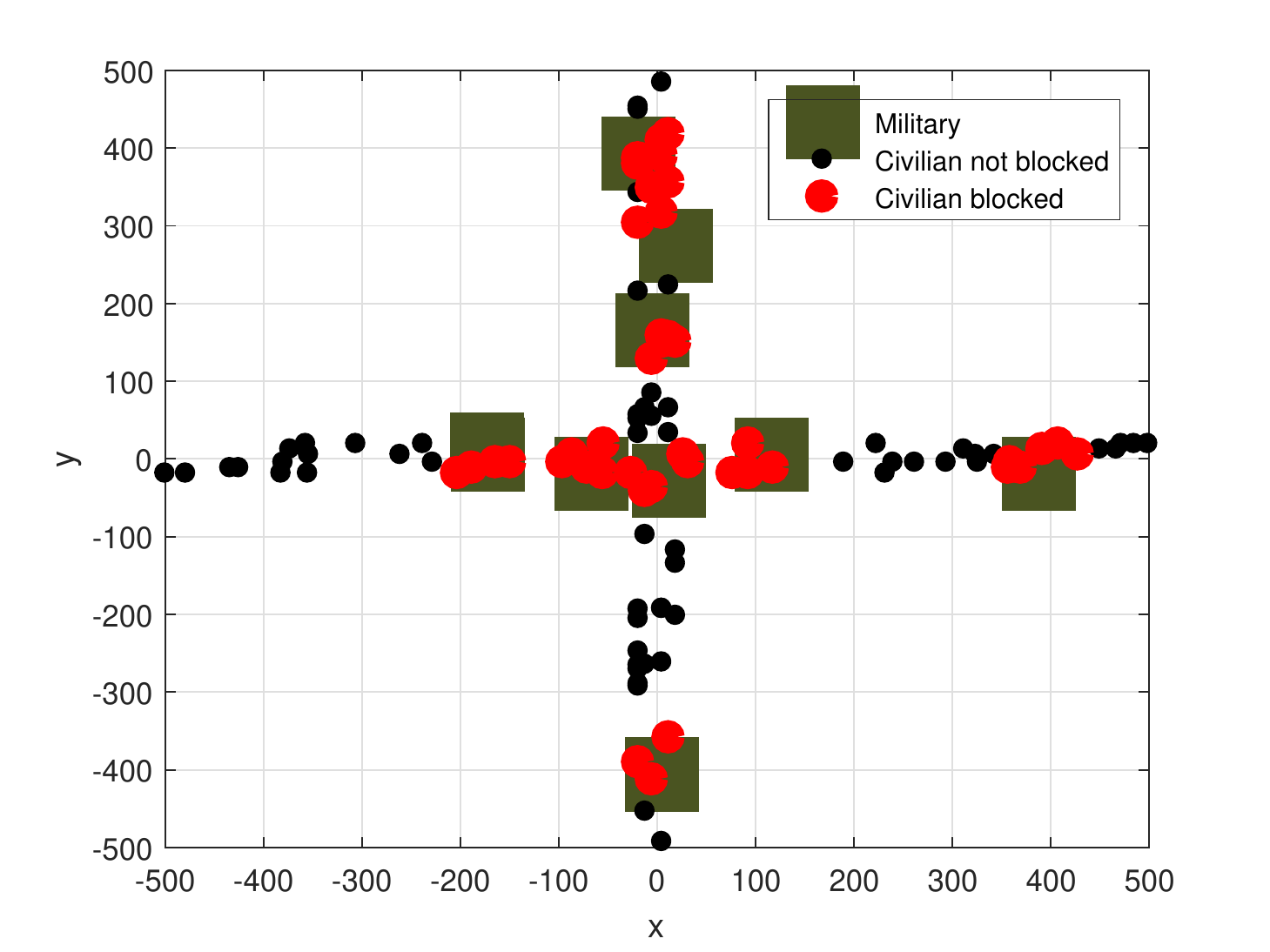}
\caption{A snapshot with $\mathsf{d}_{\text{TH}} = 50$ m}
\label{fig_snapshot_th50}
\endminipage
\minipage{0.35\textwidth}
\centering
\hspace{-0.2 in}
\includegraphics[width=\linewidth]{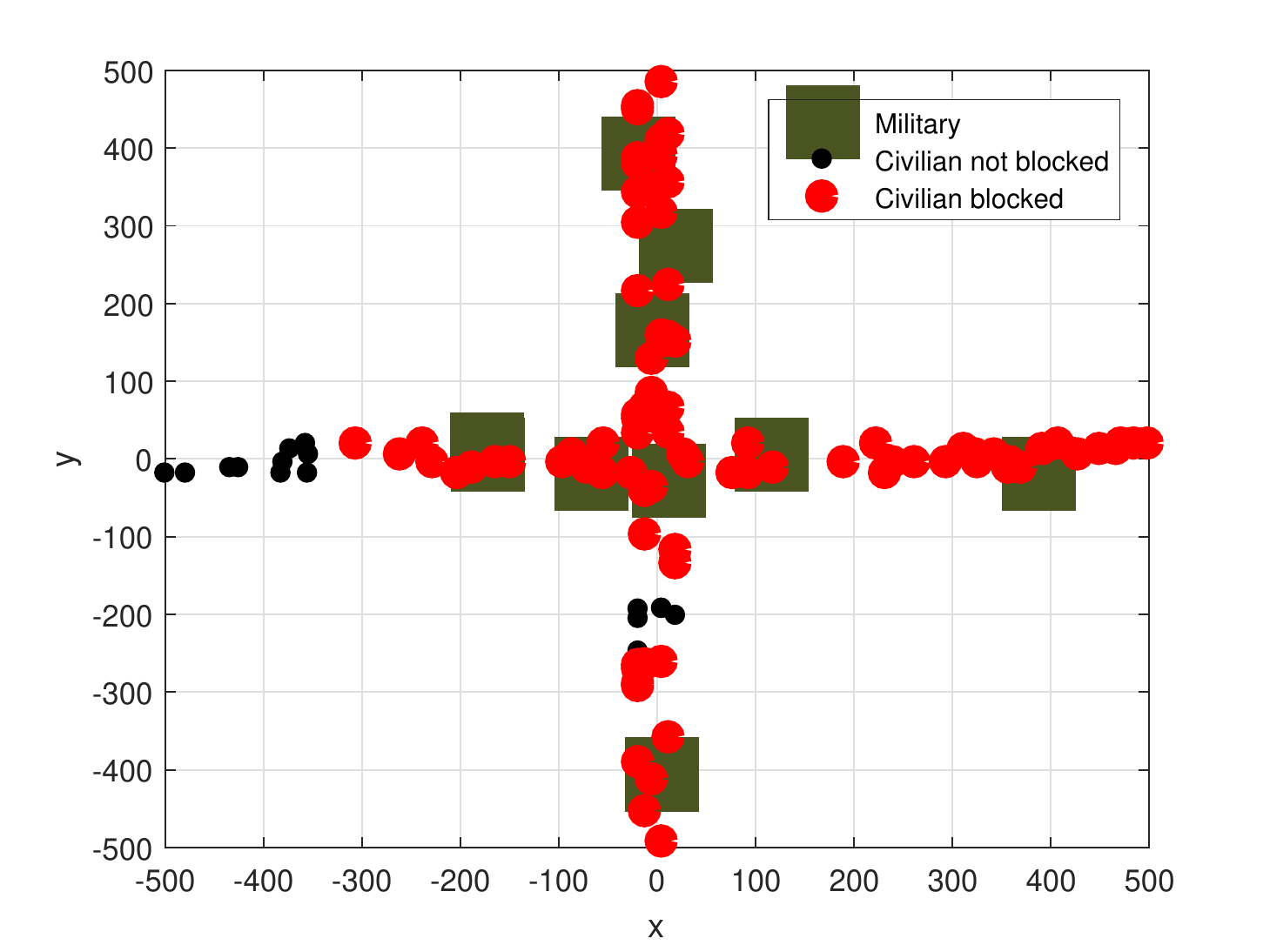}
\caption{A snapshot with $\mathsf{d}_{\text{TH}} = 150$ m}
\label{fig_snapshot_th150}
\endminipage
\minipage{0.35\textwidth}
\centering
\hspace{-0.2 in}
\includegraphics[width=\linewidth]{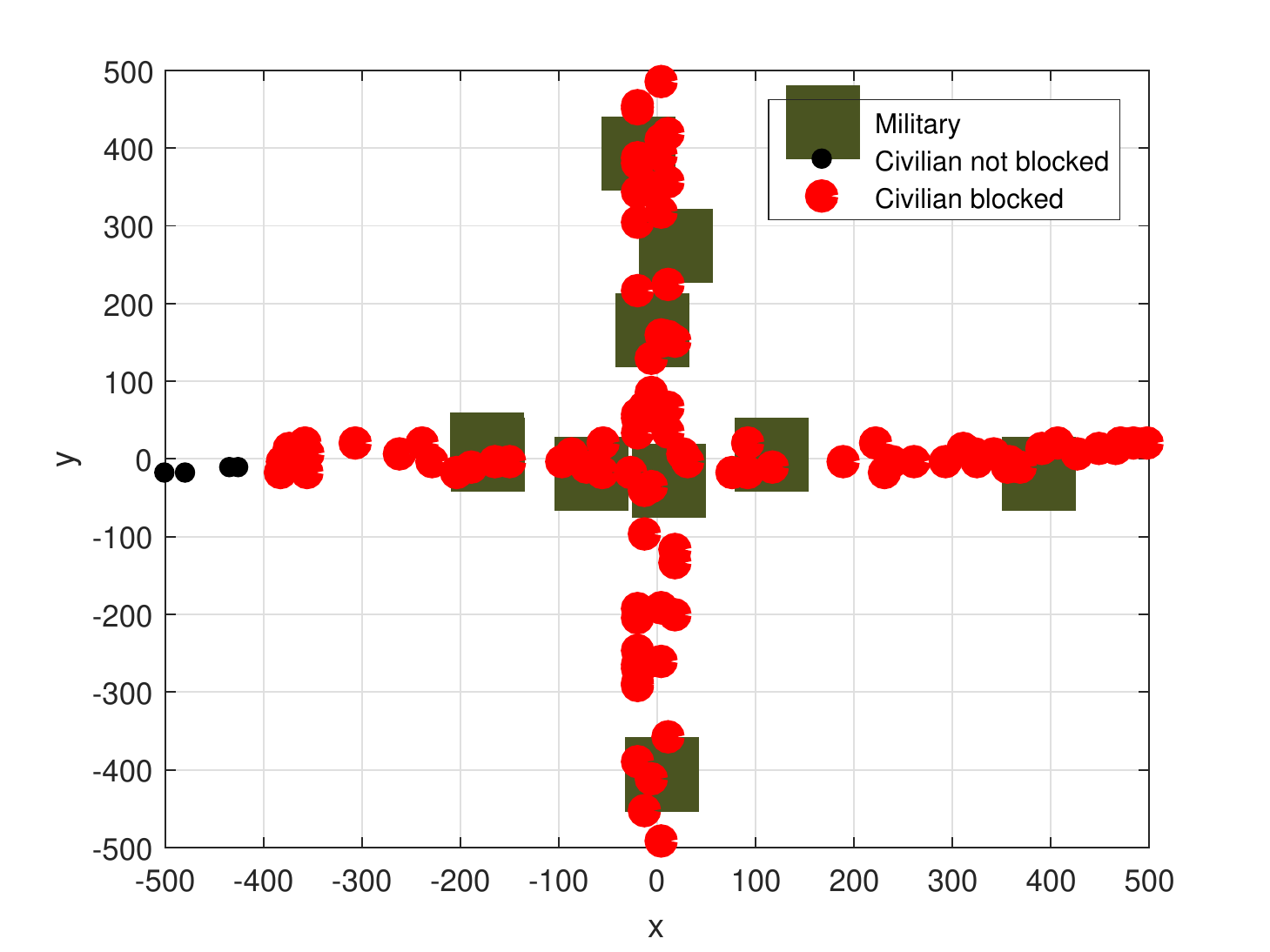}
\caption{A snapshot with $\mathsf{d}_{\text{TH}} = 250$ m}
\label{fig_snapshot_th250}
\endminipage
\vspace{-0.2 in}
\end{figure*}

%%%%%%%%%%%%%%%%%%%%%%%%%%%%%%%%%%%%%%%%%%%%%%%%%%%%%%
\section{Proposed Protocol}\label{sec_proposed}
Recall that the most significant contribution of this paper is to propose a method to improve the communications performance for military vehicles. Specifically, it proposes a protocol that filters out the data traffic generated by civilian vehicles that are too close to a military vehicle.

One important assumption is that each vehicle is able to calculate the distance to a neighboring vehicle. In fact, a variety of means are available to detect such a distance, from the Global Positioning System (GPS) to an automotive radar. The average number of sensors on a vehicle today is around 100, but that number is expected to double by 2020 as vehicles become smarter \cite{heathweb}. Based on this assumption, a civilian vehicle is capable of calculating the Euclidean distance a military vehicle within its range, which is formally written as $\mathsf{d}\left(\mathtt{x}_{m}, \mathtt{x}_{n}\right) = ||\mathbf{x_{\text{c}}} - \mathbf{x_{\text{m}}}||$.

\begin{figure}[t]
\centering
\includegraphics[width=\linewidth]{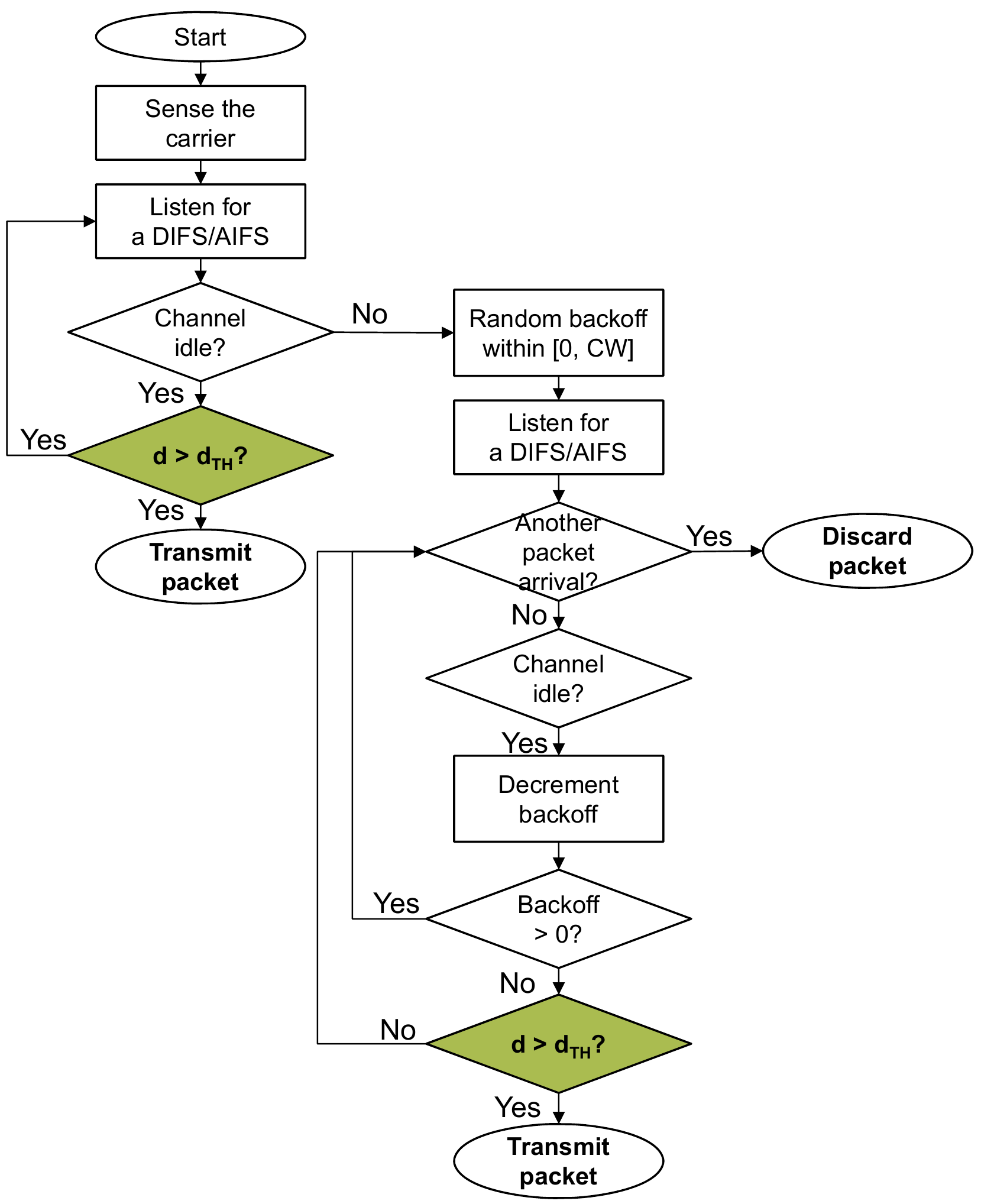}
\caption{Flowchart at a civilian vehicle with the proposed protocol}
\label{fig_flowchart}
\vspace{-0.2 in}
\end{figure}
Now, in order to prioritize the military vehicles, the proposed protocol introduces the \textit{threshold for the military-civilian distance}, which is denoted by $\mathsf{d}_{\text{TH}}$. The threshold is defined as the minimum separation distance that must be guaranteed between a civilian and a military vehicle. For instance, with $\mathsf{d}_{\text{TH}} = 100$ m, all civilian vehicles located within the 100 m range of a military vehicle will be prohibited from transmitting any packet. As such, with this protocol enabled, a military vehicle can take the priority for a channel use, which results in an improved quality of communications among military vehicles.

Fig. \ref{fig_flowchart} illustrates the flow of the proposed protocol. Notice that the proposed protocol is supposed to be applied at a civilian vehicle. Overall, it is a modification of the original IEEE 802.11 carrier-sense multiple access (CSMA) \cite{ieee80211}. Whenever a packet is about to be transmitted, the civilian vehicle must consider the distance to the closest military vehicle at the time instant, which is denoted as $\mathsf{d}$ in the flowchart. The unique distance criterion is highlighted in green color.

Figs. \ref{fig_snapshot_th50} through \ref{fig_snapshot_th250} demonstrate a snapshot with the separation distance threshold $\mathsf{d}_{\text{TH}} = \{50, 150, 250\}$ m. Note that the distribution of military and civilian vehicles are with densities of $\lambda_{m} = 10$ per $|\mathbb{R}^2|$ and $\lambda_{c} = 100$ per $|\mathbb{R}^2|$. Large `army green' squares represent the military vehicles, while small circles indicate the civilians. By applying the proposed protocol, the civilian vehicles with $\mathsf{d}(\mathbf{x}_{m}, \mathbf{x}_{c}) > \mathsf{d}_{\text{TH}}$ are blocked from transmission, which are marked as red circles. It is demonstrated that as $\mathsf{d}_{\text{TH}}$ is increased, a larger number of civilian vehicles are blocked.

\begin{figure*}[t]
\vspace{-0.2 in}
\centering
\minipage{0.35\textwidth}
\hspace{-0.2 in}
\includegraphics[width=\linewidth]{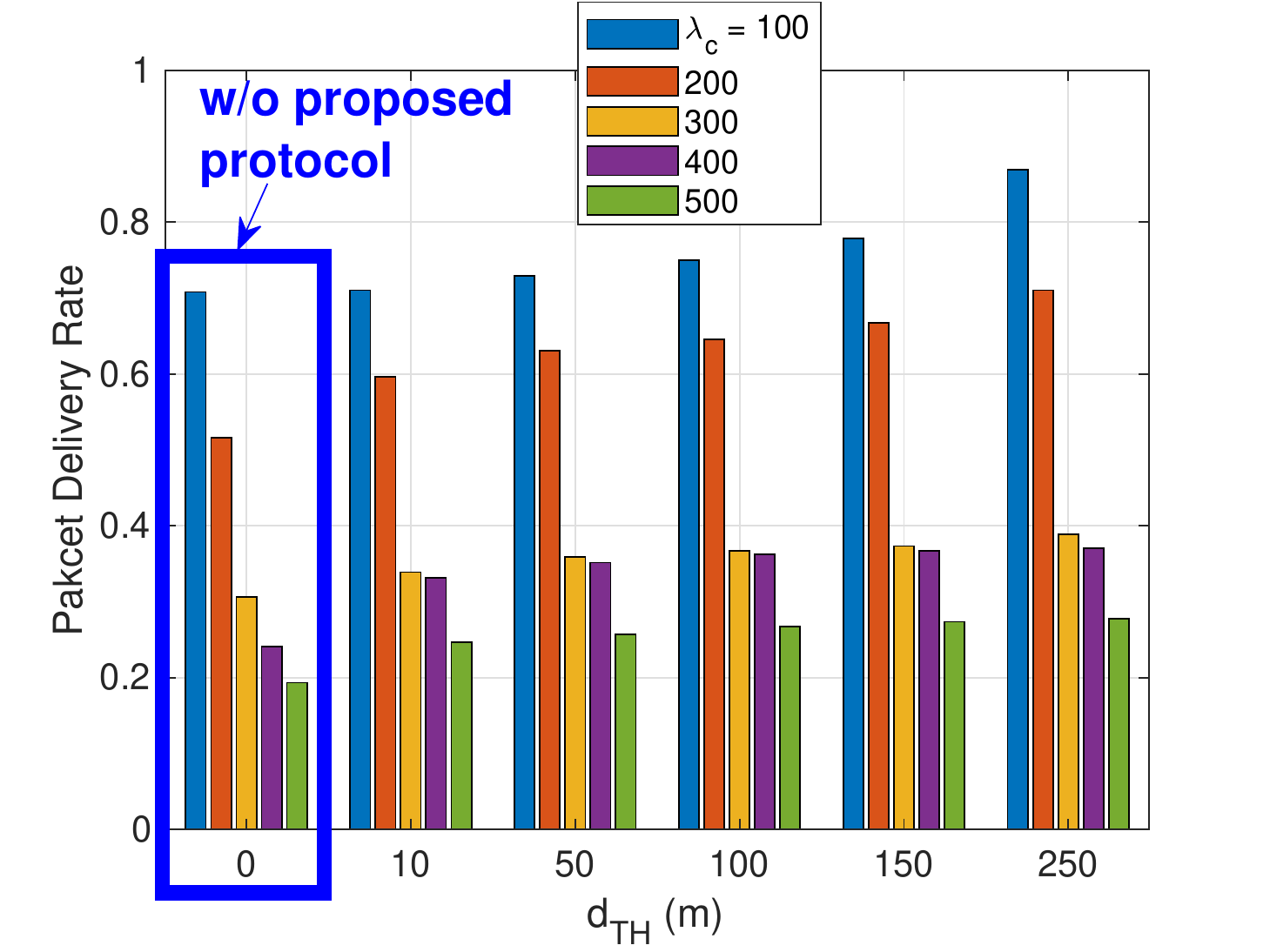}
\caption{PDR versus $\mathsf{d}_{\text{TH}}$}
\label{fig_ps}
\endminipage
\minipage{0.35\textwidth}
\centering
\hspace{-0.2 in}
\includegraphics[width=\linewidth]{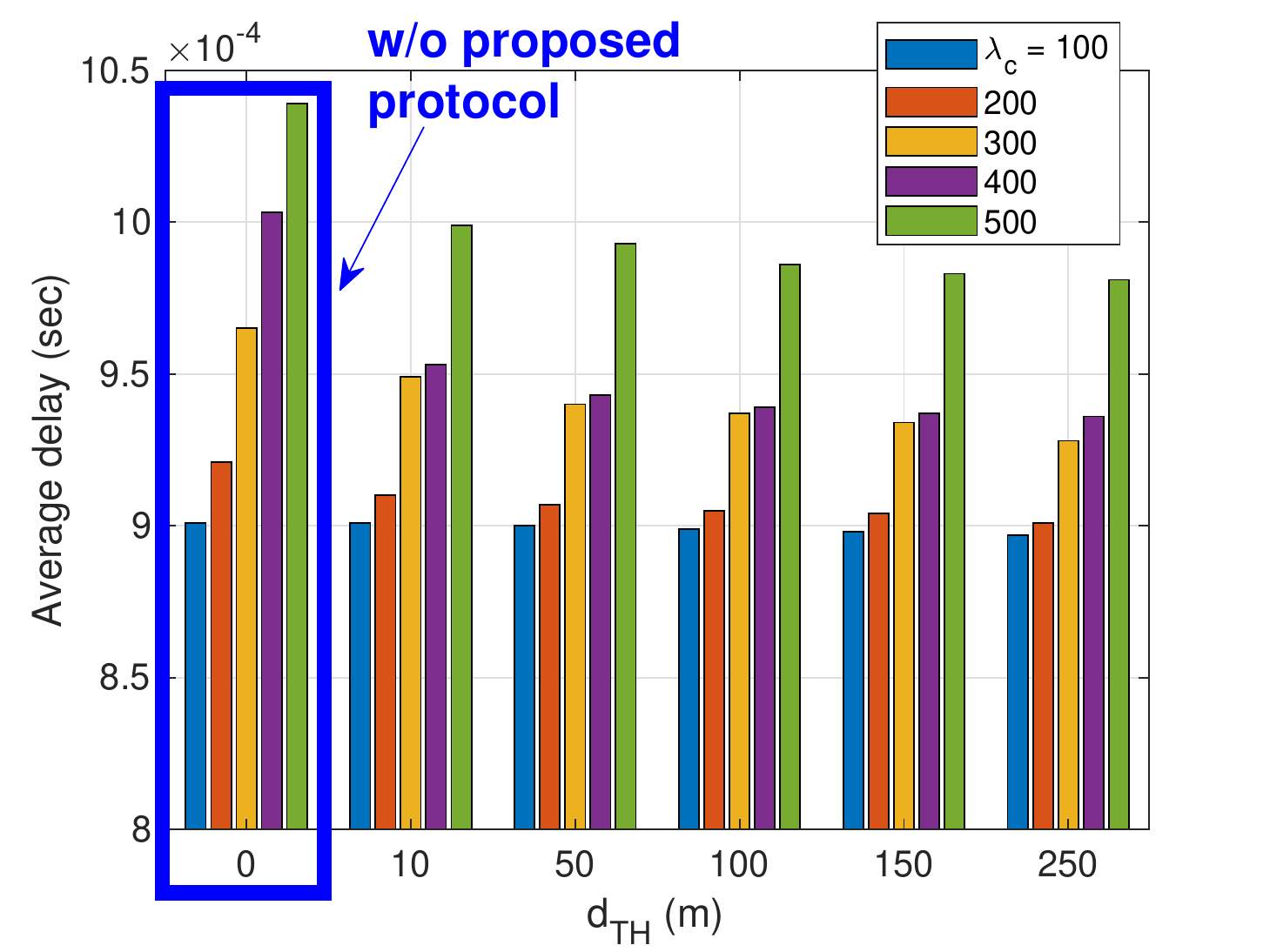}
\caption{Average delay versus $\mathsf{d}_{\text{TH}}$}
\label{fig_delay}
\endminipage
\minipage{0.35\textwidth}
\centering
\hspace{-0.2 in}
\includegraphics[width=\linewidth]{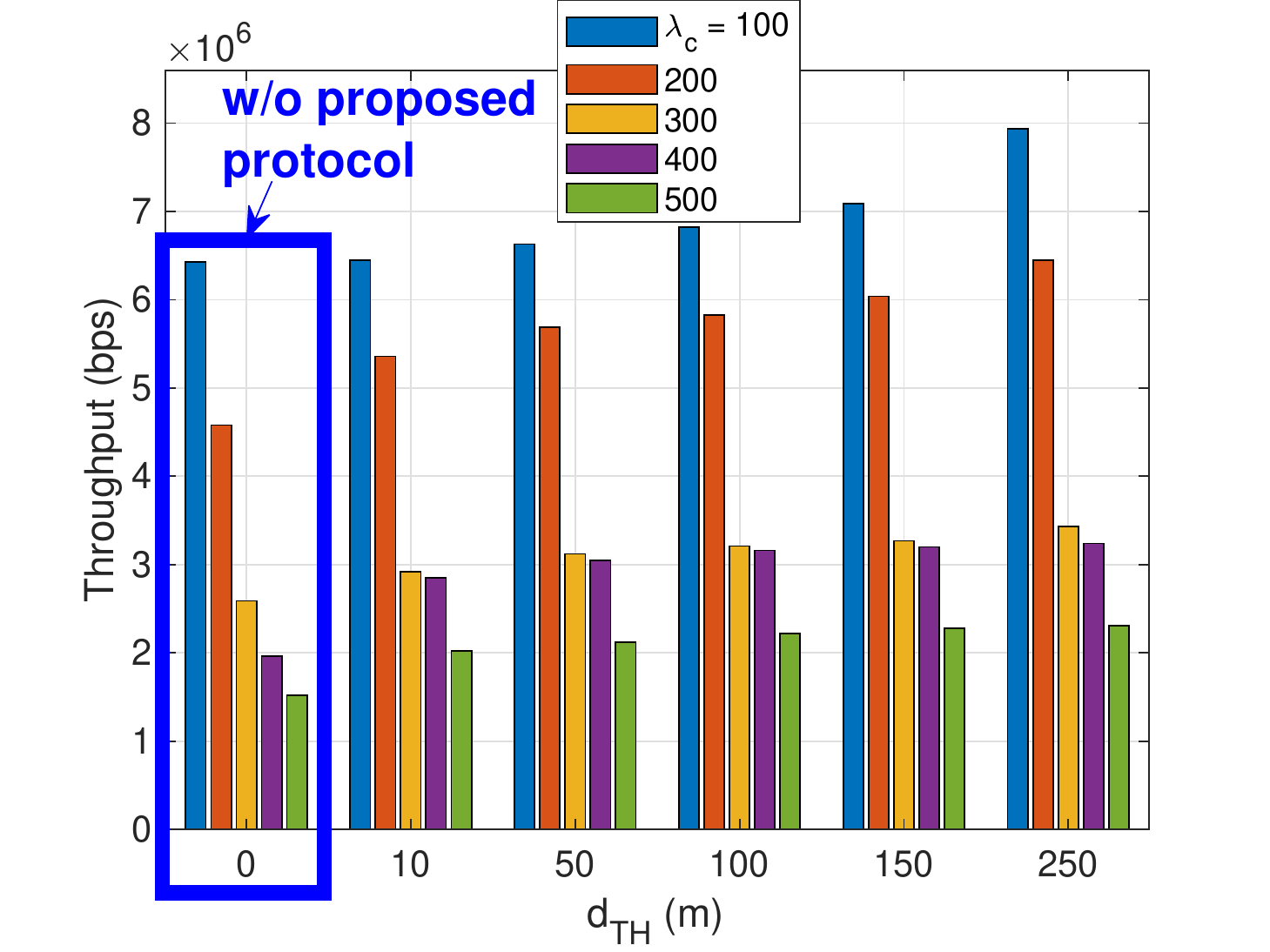}
\caption{Throughput versus $\mathsf{d}_{\text{TH}}$}
\label{fig_throughput}
\endminipage
\vspace{-0.1 in}
\end{figure*}

\begin{table}[t]
\caption{Parameters for simulations}
\centering
\begin{tabular}{p{0.5in} p{2 in}}
\hline
\textbf{Parameter} & \textbf{Value}\\
\hline\hline
Bandwidth& 10 MHz\\
BSM rate & 10 Hz\\
$CW$& 15\\
$\lambda_{\text{m}}$ & 10 per $||\mathbb{R}^2||$\\
DIFS& 64 $\mu$s\\
SIFS& 32 $\mu$s\\
$\delta$ & 1 $\mu$s\\
$r_{tx}$& 500 m\\
$T_{slot}$ & 13 $\mu$s\\
$T_{pl}$ & 1023 bytes\\
$T_{pl}$& 8184 bits \cite{alkadeki2015performance} ${\rm{~}}$ (8184$\times$10$^{-7}$ = 814.8 $\mu$s)\\
$T_{c}$& 8713 bits \cite{bianchi2000performance} ${\rm{~}}$ (8713$\times$10$^{-7}$ = 871.3 $\mu$s)\\
$T_{s}$& 8982 bits \cite{bianchi2000performance} ${\rm{~}}$ (8982$\times$10$^{-7}$ = 898.2 $\mu$s)\\
\hline
\end{tabular}
\vspace{-0.2 in}
\label{table_parameters}
\end{table}

%%%%%%%%%%%%%%%%%%%%%%%%%%%%%%%%%%%%%%%%%%%
\section{Performance Evaluation}
This section presents simulation results to evaluate the performance of the proposed protocol. The parameters are listed in Table \ref{table_parameters}. Considering the large number of parameters and wider selection possibility for their values, the values chosen in Table \ref{table_parameters} are examples to show that the proposed protocol is effective. For instance, this simulation sets CW to 15 while it can have the range of \{15, 31, 63, 127, 255, 511, 1023\} in DSRC \cite{usc08}. However, the simulation framework is general and thus can easily be extended to any other values for all the parameters.

Simulations were performed via MATLAB. Military and civilian vehicles were distributed on the two-dimensional space $\mathbb{R}^2$ illustrated in Fig. \ref{fig_system_model} with the densities of $\lambda_{m} = 10$ per $|\mathbb{R}^2|$ and $\lambda_{c} = \{100, 200, 300, 400, 500\}$ per $|\mathbb{R}^2|$, respectively. The performance of the proposed protocol is evaluated based on the three metrics described in Section \ref{sec_analysis_metrics}--namely PDR, average delay, and throughput.

Recall that the threshold for military-civilian separation distance, $\mathsf{d}_{\text{TH}}$, is the parameter that determines the level of operation of the proposed protocol. As such, for all the results given in Figs. \ref{fig_ps} through \ref{fig_throughput}, $\mathsf{d}_{\text{TH}}$ forms the X axis to provide comparisons. Notice that $\mathsf{d}_{\text{TH}} = 0$ means ``no application of the proposed protocol.'' It provides a direct insight how the proposed protocol improves the communications performance of a military vehicle.

Fig. \ref{fig_ps} shows PDR, or the probability of a successful packet delivery, which is given in Eq. (\ref{eq_ps}). One observe that a higher PDR is achieved with a larger value for $\mathsf{d}_{\text{TH}}$. The rationale is that an increase in $\mathsf{d}_{\text{TH}}$ means suppression of a larger number of civilian vehicles from BSM transmissions, which as a consequence gives military vehicles a higher chance to win a channel.

Fig. \ref{fig_delay} evaluates the proposed protocol in terms of the average delay, which is formulated in Eq. (\ref{eq_Ttot}). Similarly with the discussion on PDR, application of this protocol reduces the average delay as it decreases more civilian vehicles competing a channel with military vehicles. The resulting average delay ranges from 0.9 msec to 1 msec remains far below the ``allowable latency,'' which is nominally understood to be 100 msec \cite{nhtsa}.

Fig. \ref{fig_throughput} compares the performance of the proposed protocol according to $\mathsf{d}_{\text{TH}}$. Selection of a larger value for $\mathsf{d}_{\text{TH}}$ leads to a higher throughput, since it allows a military vehicle to transmit a larger number of packets within a unit time.

%%%%%%%%%%%%%%%%%%%%%%%%%%%%%%%%%%%%%%%%%%%%%%%%%%%%%
\section{Conclusions}
This paper addressed the channel congestion for a DSRC system operating in a mixture of military and civilian vehicles. A novel protocol was proposed that increases the chance of a successful packet delivery for a military vehicle. The protocol is designed to prioritize the military vehicles in competition for a channel. It prohibits a transmission from a civilian vehicle with the distance to the closest military vehicle being smaller than a threshold. The simulations showed the proposed protocol's performance based on the three widely accepted metrics--PDR, average delay, and throughput. According to the simulation results, the protocol effectively increased the performance of communications at a military vehicle with respect to all the three metrics.

Multiple key design insights for the military communications were drawn from this paper's results. Specifically, they revealed the required separation distances between a civilian and a military vehicle in terms of three different metrics. Furthermore, they showed the impact of parameters on the performance of DSRC for military's urban operations--\textit{i.e.}, the threshold distance between a military and a civilian, and the traffic density of civilian vehicles in relation to the military.

As such, this work has many possible extensions. For instance, in this paper, we considered only the DSRC-based communications. One possible extension is to incorporate discussions on coexistence of military and civilian in a cellular vehicle-to-everything (C-V2X) system as well. An even more promising avenue of future work is to develop a comprehensive algorithm to enable coexistence under a mixture of DSRC and C-V2X in the 5.9 GHz band.

%%%%%%%%%%%%%%%%%%%%%%%%%%%%%%%%%%%%%%%%%%%%%%%%%%%%%%%%

\end{document}